\documentstyle[aps,twocolumn,floats,prl]{revtex}

\newcommand{\beq}{\begin{equation}}
\newcommand{\eeq}{\end{equation}}
\def\beqa{\begin{eqnarray}}
\def\eeqa{\end{eqnarray}}
\def\p{\partial}
\def\mp{m_{\rm pl}}
\def\lap{\lower.5ex\hbox{$\; \buildrel < \over \sim \;$}}
\def\gap{\lower.5ex\hbox{$\; \buildrel > \over \sim \;$}}

\input psfig

\begin{document}
\title{Quintessential Inflation}
\author{P. J. E. Peebles\\
Joseph Henry Laboratories, Princeton University, and \\
Institute for Advanced Study, Princeton NJ 08544\\
and\\
A. Vilenkin\\
Department of Physics\\
Tufts University, Medford MA 02155}
\date{\today}
\maketitle
\begin{abstract}
We present an explicit observationally acceptable model for evolution from 
inflation to the present epoch under the assumption that the entropy
and matter of the familiar universe are from gravitational particle
production at the end of inflation.  This  
eliminates the problem of finding a satisfactory coupling of the
inflaton and matter fields. Since the inflaton potential $V(\phi)$ may
be a monotonic function of the inflaton $\phi$, the inflaton energy
could produce an observationally  significant effective cosmological
constant, as in quintessence.
\end{abstract}

\section{Introduction}

A satisfactory inflation model for the very early universe has to
account for the entropy in matter --- the fields of
present-day physics or their predecessors. The usual assumption is that the 
entropy 
comes from the decay of  
the inflaton field or fields $\phi$ whose stress-energy tensor drives inflation. 
In 
this 
scenario the inflaton rolls toward a minimum of its 
potential where it oscillates, and the oscillations are damped by the production 
of 
quanta of fields coupled to $\phi$. The couplings have to be extremely weak, to protect the flat inflaton potential which is necessary for a successful 
inflation 
picture, 
but at the same time strong enough to allow efficient thermalization of the 
inflaton 
energy density. Models satisfying these criteria do not arise naturally in particle 
physics; they have to be constructed {\it ad hoc}. 

A less commonly discussed alternative is that the entropy in the
matter fields comes from gravitational particle production at the end
of inflation.\footnote{In another terminology, matter fields end up in
squeezed states after inflation \cite{Grish}.} Since the inflaton does not 
interact
with matter, and its energy density can roll monotonically toward
zero, it is a candidate for a present-day effective cosmological
constant. This picture merits broader discussion. To this end we
present a specific example that we believe satisfies the main
observational constraints. Our choice of the inflaton potential is
presented in the next section. The results of a numerical integration
of the evolution from inflation to the present are shown in Figure~1.

Ford\cite{Ford} first investigated gravitational particle production at the end of 
inflation, and pointed out that
particles created this way could account for the entropy
of the present universe. Inflation would be followed by a period of
expansion dominated by the energy density $\rho _\phi$ of the
inflaton. Ford noted that if the kinetic part $\dot\phi ^2/2$ were
dominant the energy density would decay with the expansion factor
$a(t)$ as $\rho _\phi\propto a^{-6}$, so the expansion could become
radiation-dominated before light element
production. Spokoiny\cite{Spok} added the idea that the inflaton could
end up as the scalar field that Peebles and Ratra\cite{PR} proposed
serves as an effective present-day cosmological
constant. Spokoiny\cite{Spok} presented examples of potentials $V(\phi
)$ in which there is a transition from inflation to kinetic
energy-dominated expansion, and he considered the conditions under
which the expansion later is  dominated again by the inflaton. Since
gravity waves are produced by inflation with the same energy density
per component as the matter fields, the entropy at very high redshift
must be distributed among 
enough matter field components that gravitational waves do not make a
significant contribution to the expansion rate during light element
production. We discuss this condition in \S VI, along with the
possible detection of the gravitational waves by future
detectors. Other possible connections to observations include the
effect of the kinetic energy-dominated expansion on relict particle
abundances, as discussed by Kamionkowski and Turner\cite{KT}, and on
baryogenesis, as considered by Joyce \cite{Joyce1} and Joyce and
Prokopec\cite{Joyce2}. There may also be a conceptual value to the
postulate that the inflaton potential resides in a sector that is not
coupled to matter fields, and so may assume a simple and possibly
``natural''  form.

Terminology might be mentioned. Spokoiny\cite{Spok}
calls the kinetic energy-dominated period ``deflationary;'' 
Joyce\cite{Joyce1} prefers ``kination.'' Caldwell, Dave, and 
Steinhardt\cite{QUINT} 
propose the name ``quintessence'' for  a field that acts like Einstein's 
cosmological 
constant $\Lambda$. Since the model under discussion reduces the role of the 
inflaton
to the essential operation of driving inflation, and adds the possibility that 
the 
inflaton ends up as quintessence, we call the picture  ``quintessential 
inflation.''

\section{The Model}

We consider inflation driven by a single scalar
inflaton field $\phi$ that interacts only with gravity and itself
by the potential $V(\phi)$. The matter fields are scalar, spinor and gauge 
fields of 
some grand unified theory, and are supposed to be in their vacuum states during 
inflation.

The homogeneous part of the inflaton field satisfies
\beq
{d^2\phi\over dt^2} + 3{\dot a\over a}{d\phi\over dt} = -{dV\over d\phi},
\label{1}
\eeq
where the expansion rate is
\beq
H^2 = (\dot a/a)^2 = (\rho_\phi + \rho_m)/\mp^2,
\label{2}
\eeq
and the Planck mass is written as  
\beq
\mp = (8\pi G/3)^{-1/2} = 4.2\times 10^{18}\hbox{ GeV},
\label{3}
\eeq
with $\hbar = 1 = c$. The energy density in the inflaton is
\beq
\rho_\phi = \dot\phi ^2/2 + V(\phi ),
\label{4}
\eeq
and $\rho_m(t)$ is the matter energy density after inflation.

We adopt the inflaton potential 
\beqa
V &=& \lambda (\phi^4 + M^4)~~~~~{\rm for}~~\phi < 0,\nonumber \\
 &=& {\lambda M^8\over{\phi^4 + M^4}}~~~~~{\rm for}~~ \phi\geq 0,
\label{5}
\eeqa
At $-\phi\gg M$
this is a ``chaotic" inflation potential \cite{Linde}; at
$\phi\gg M$ it is a ``quintessence" form of the type
considered in Refs.\cite{PR}, \cite{QUINT}, \cite{QUINT2}. We adopt
\beq
\lambda = 1\times 10^{-14},
\label{6}
\eeq
from the condition that present-day large-scale structure grows from
quantum fluctuations frozen into $\phi$ during inflation.\footnote{On
scales of astronomical interest the only difference from ``chaotic''
inflation with $V(\phi)=\lambda\phi^4$ is the increased
expansion factor to the end of inflation. Since the adibatic density
fluctuations from the frozen inflaton depend on the logarithm of the
expansion factor \cite{Linde}, the increased expansion is not
significant. In conventional inflation the value of $\lambda$ from the
COBE normalization   
\cite{lambda} is three times the rounded value in Eq. (\ref{6}). Again, the 
precise value of $\lambda$ is not important.}  The
following sections establish bounds on the constant energy parameter
$M$. The example in Figure~1 assumes the present density parameter in
matter is $\Omega _m\sim 0.3$, with $\Omega _\phi = 1-\Omega _m\sim
0.7$ in the inflaton. Then with the model for matter and radiation in
\S\S V and VI,
\beq
M = 8\times 10^5\hbox{ GeV}.
\label{6a}
\eeq

\section{Kinetic Energy-Dominated Expansion}

We assume the inflaton starts at $\phi\ll -\mp$ and rolls toward zero. Inflation 
ends 
at $\phi\sim-\mp$ when a substantial part of the potential energy has turned 
into the 
kinetic energy of the inflaton. This is followed by a kinetic energy-dominated 
peri
od where $V(\phi )$ and the matter energy
density terms in Eqs. (\ref{1}) to (\ref{4}) are negligible. Here the solution 
to Eq. 
(\ref{1}) is ${\dot\phi}\propto a^{-3}$,
and the energy density is
\begin{equation}
\rho_\phi\approx{\dot\phi}^2/2\sim\lambda\mp^4(a_x/a)^6,
\label{7}
\end{equation}
where $a_x$ is the expansion factor at the end of inflation. This expression in 
the 
expansion rate Eq. (\ref{2}) yields
\begin{equation}
a\sim t^{1/3},~~~~~~\phi = 2^{1/2}\mp\ln(a/a_x).
\label{8}
\end{equation}
The central point, as shown by Ford\cite{Ford} and Spokoiny\cite{Spok}, is that 
the 
inflaton energy density decreases faster than that of radiation, so if the 
potential 
remains negligible the energy of the particles produced at the end of inflation 
eventu
ally dominates. Our potential decreases only logarithmically with $a$,
$V\sim\lambda M^8\mp^{-4}\left[\ln(a/a_x)\right]^{-4}$,
so if the radiation remains subdominant $V(\phi)$ approaches the kinetic energy 
density at scale factor $a_*$ given by
\begin{equation}
{a_*\over a_x}\sim\left({\mp\over M}\right)^{4/3}\ln^{2/3}\left({\mp\over 
M}\right).
\label{10}
\end{equation}
The expansion then enters inflation from which it
never recovers. Our model has a chance to work\footnote{One sees from Eqs. (\protect\ref{7}) and (\ref{8}) that $\rho _\phi$ after inflation remains 
dominated by 
the kinetic energy part until $\rho _m$ becomes important if the potential 
varies as 
rapidly 
as $V\propto e^{-\phi/\phi_c}$ with $\phi _c < m_{\rm pl}/(3\sqrt{2})$.
In our example the exponential variation of $V$ with $\phi$ is
replaced by the rapid functional change of $V$ at $\phi\sim 0$ because
of the small value of $M/\mp$.} only if the matter energy density
$\rho_m(a_x)$ at the end of inflation is large enough to dominate the
inflaton energy density at $a=a_r < a_*$. We now turn to the estimate
of $\rho_m$.

\section{Particle Creation and the Radiation Era}

Particle creation at the end of inflation can be studied by
the standard methods of quantum field theory in curved spacetime
\cite{BD}. For a  massless scalar field $\chi$ described by the
Lagrangian 
\begin{equation}
L = {1\over 2}(\p_\mu\chi)^2 - {1\over 2}\xi{\cal R}\chi^2,
\label{11}
\end{equation}
where ${\cal R}$ is the scalar curvature, this has been done by
Ford \cite{Ford} in the limit of nearly conformal coupling, 
$|\xi - {1\over 6}|\ll 1$. He 
assumes that during inflation the spacetime is close to de
Sitter, and that inflation is immediately followed by radiation-dominated 
expansion. 
His result for the energy density of the created particles is 
\beq
\rho_m = RH_x^4(a_x/a)^4,
\label{12}
\eeq
where $H_x$ is the Hubble parameter, inflation ends at $a=a_x$, and the 
numerical 
factor is\footnote{The
``sudden" approximation, in which the de Sitter line element is
matched to the radiation-dominated one, is adequate for 
modes with wave numbers $k\ll H_x$ and can be expected to give
correct orders of magnitude for $k\sim H_x$. Particle production
is strongly  suppressed for $k\gg H_x$. The main contribution to
$\rho_m$ is given by the modes  with $k\sim H_x$, and thus Eq.
(\ref{13}) can be regarded only as an
order-of-magnitude estimate.}
\beq
R\sim 10^{-2}(1 - 6\xi)^2.
\label{13}
\eeq
Ford's analysis \cite{Ford} was extended to arbitrary values of
$\xi$ \cite{DV} and to arbitrary power-law expansion after
inflation \cite{MG}. In all cases $R\sim 10^{-2}$. We assume minimal coupling, 
$\xi = 
0$, and adopt $R=0.01$ per scalar field component. 

Free massless spinor and gauge fields are 
conformally-invariant, so their contribution to $\rho_m$ is suppressed relative 
to 
scalar fields.\footnote{In the case of interacting fields, particle production 
can 
occur  
even if the field equations are conformally invariant. This is
due to the  conformal anomaly of the quantum theory \cite{BD2}.}
Thus in a simple model of gravitational production of quanta with negligible 
rest mass 
the matter energy density is given by Eq. (\ref{12}) with
\beq
R\sim 10^{-2}N_s,
\label{14}
\eeq
where $N_s$ is the number of scalar fields.

We consider next the thermalization of $\rho _m$. 
For $\lambda = 10^{-14}$ the Hubble
parameter at the end of inflation is 
\beq
H_x\sim\lambda^{1/2}\mp\sim 10^{12}\hbox{ GeV},
\label{15}
\eeq
at the earliest filled squares in Figure~1. 
The created particles have typical energy $\epsilon\sim H_x(a_x/a)$ and number 
density 
$n\sim R\epsilon^3$. As  
Spokoiny\cite{Spok} noted, the particles interact by the exchange of gauge 
bosons and 
establish thermal equilibrium among the fermions and gauge bosons when the 
interaction 
rate $n\sigma$ becomes comparable to the expansion rate $H$. Here, 
$\sigma\sim\alpha^2\epsilon^{-2}$ and $\alpha\sim 0.1 - 0.01$ is the gauge coupling constant. 
With 
$H\sim  H_x(a_x/a)^3$, thermalization occurs at 
\beq
a_{\rm th}/a_x\sim\alpha^{-1}R^{-1/2}\sim(10^2 - 10^3)N_s^{-1/2},
\label{17}
\eeq
only a few orders of magnitude expansion from the end of inflation, at 
temperature
\beq
T_{\rm th}\sim\rho_m^{1/4}(a_{\rm th})\sim R^{3/4}\alpha 
	H_x\sim 10^9N_s^{3/4}\hbox{ GeV}.
\label{18}
\eeq

During kinetic-energy dominated expansion the ratio of energy
densities in matter and the inflaton is (Eqs. \ref{7} and \ref{12})  \beq
\rho_m/\rho_\phi\sim\lambda R(a/a_x)^2.
\label{19}
\eeq
Radiation-dominated expansion thus begins at expansion factor 
\beq
a_r/a_x\sim (\lambda R)^{-1/2},
\label{20}
\eeq
at the second earliest filled squares in Figure~1. 
We require $a_r<a_*$, where $a_*$ is given by Eq.(\ref{10}).  With $R\sim 
0.01-1$, 
this gives the condition 
\beq
M<(10^{13} - 10^{14})\hbox{ GeV},
\label{20a}
\eeq
well above the value for quintessence (Eq. \ref{6a}). The temperature at the 
start of 
radiation-dominated
expansion is, from Eq. (\ref{14}), 
\beq
T_r\sim \rho_m^{1/4}(a_r) \sim \lambda R^{3/4}\mp 
\sim 10^3N_s^{3/4}\hbox{ GeV}.
\label{22}
\eeq

\section{Evolution Through the Radiation Era to the Present}

At $a\sim a_r$ the inflaton satisfies 
\beq
\phi_r\sim 2^{1/2}\mp\ln (\lambda R)^{-1/2},  \qquad
\dot\phi_r\sim R^{3/2}\lambda^2\mp^2.
\label{24}
\eeq
In an extended period after $a_r$ the potential gradient term in the field 
equation 
(\ref{1}) for $\phi$ is
negligible and the solution for radiation-dominated expansion with the initial 
conditions (\ref{24}) is
$\phi = A - B (t_r/t)^{1/2}$, with $A\approx \phi_r$ and $B\sim m_{pl}$. That 
is, 
$\phi$ 
is increasing quite slowly.
at
During matter-dominated expansion there is a stable increasing solution, 
\beq
\phi = k\lambda^{1/6}M^{4/3}t^{1/3},
\label{28}
\eeq
where $k = (72/5)^{1/6}$ in the radiation era and $k = 3^{1/3}$
in the nonrelativistic matter era. The inflaton remains close to constant at 
$\phi\sim\phi _r$ until this solution becomes comparable to $\phi _r$, at 
\beq
t_a \sim  \lambda^{-1/2} m_{pl}^3M^{-4} [\ln (\lambda R)^{-1/2}]^3 
\sim 10^{11}\hbox{ y},
\label{28a}
\eeq
for our parameters. Since this is much greater than the present expansion time 
$\rho 
_\phi$ is nearly constant from the end of  kinetic-energy dominated expansion to 
the 
present, as in Figure~1. 
The present energy density in the inflaton thus is $\rho _\phi\sim
V(\phi _r)$, and the condition that this is comparable to the total
present energy density is 
\beq
M\sim\lambda ^{-1/8}\mp ^{3/4}H_o^{1/4}[\ln (\lambda R)^{-1}]^{1/2}
\sim 10^6\hbox{ GeV},
\label{27a}
\eeq
close to the numerical result (Eq. \ref{6a}).

When the energy density $\rho _\phi$ in the inflaton remains subdominant to 
$a>a_r$ the expansion factor from the end of inflation to the time when $\rho 
_\phi$ reaches the nearly constant value $V(\phi _r)$ follows as in Eq. 
(\ref{10}), 
\begin{equation}
a_*/a_x\sim (\mp\phi _r)^{2/3}/M^{4/3}.
\label{10a}
\end{equation}
This is $a_*/a_x\sim 10^{18}$ for our parameters.

It might be noted that the model is not sensitive to the power law index in the 
potential after inflation. The potential in Eq. (\ref{5}) at $\phi > 0$ may be 
generalized to 
\beq
V = {\lambda M^4\over1 + (\phi /M)^\alpha }.
\label{5a}
\eeq
Since $V$ at $a_* < a < a_o$ is determined by the density parameter
$\Omega _\phi = 1-\Omega _m =0.7$ and Hubble's constant, which fix the present 
value 
of $\rho _\phi$, the combination $\lambda M^{4 + \alpha }/\phi _r^\alpha$ is 
independent of $\alpha$. For $\alpha = 4$ and $\lambda = 1\times 10^{-14}$ we 
require 
$M/\mp = 2\times 10^{-13}$ to agree with $\Omega _\phi$. For the same value of 
$\lambda$ and  $\alpha = 6$ we need $M/\mp = 1.3\times 10^{-10}$, and, for 
$\alpha = 2$, $M/\mp = 4\times 10^{-18}$. For this range of values of the power law index $\alpha$ and the indicated scaling of $M$ the evolution to the present is very similar to 
what is shown in Figure~1. 

At $a>a_0$, when the inflaton energy dominates, the inflationary
expansion can be described in the slow roll approximation, which neglects  
$\ddot \phi$ in Eq. (\ref{1}) and the kinetic energy density term in Eq. (\ref{4}). This gives
\beqa
\phi &\approx &\phi_r (1+CH_0 t)^{1/4}, \nonumber \\
a &\propto &\exp\left[{2\over{C}}(1+CH_0 t)^{1/2}\right],
\eeqa
where $C=16m_{pl}^2/3\phi_r^2\approx 0.01$ 
and we have used $H_o\approx V(\phi_r)/m_{pl}^2$ for the present 
expansion rate. The expansion rate remains nearly constant
and the growth of $\phi$ is approximately linear in $t$
until the time $t_b \sim 1/CH_o \sim 10^{12}$~y.

\section{Gravitational waves}

Overproduction of gravitational waves is one of the dangers to be
watched for in quintessential inflation. Gravitons are described by the same 
equation 
as a minimally coupled
scalar field, so the energy density of gravitons at the end of
inflation is twice that of a single scalar field because of the two graviton 
polarization states \cite{Allen}. Quanta of the inflaton field
$\phi$ will also be produced, so the ratio of energy densities
in the graviton-inflaton quanta and the matter density (Eqs. \ref{12}, \ref{14}) 
at 
the end of inflation is 
\beq
r_x={\rho_{gi} \over \rho_m}(a\sim a_x) \sim {3 \over N_s}.
\label{34}
\eeq
The annihilation of field quanta as the temperature falls below
the mass conserves entropy (apart from the neutrinos at $z\sim 10^{10}$), so the 
matter
temperature varies as $T_m\propto a^{-1}{\cal N}^{-1/3}$, where 
${\cal N}(T)$ is the effective number of spin degrees of freedom
in equilibrium at temperature $T$. Thus the mass
fraction in graviton and inflaton quanta immediately before electron-positron 
annihilation is 
\beq
r_n={\rho_{gi} \over \rho_m}(a\sim a_n) \approx {3 \over N_s}\left(
{{\cal N}_n \over {\cal N}_{\rm th}} \right)^{1/3}< 0.07\, . 
\label{35}
\eeq
At matter thermalization following the end of inflation, 
${\cal N}_{\rm th}\sim 10^2-10^3$ . When light element
production commences at $z\sim 10^{10}$, 
${\cal N}_n=10.75$ (in radiation
with ${\cal N}=2$, electron-positron pairs, and three families
of neutrinos). One extra
neutrino family would increase the mass density by 16\% . As 
indicated in Eq. (\ref{35}), the standard model for the light elements
requires that the gravitons add significantly less than this \cite{Allen}.

In a minimal GUT, the only light particles with masses below
$H_x$ are those of the standard model. Here $N_s=4$ (the electroweak Higgs 
doublet), 
${\cal N}_{th}\sim 10^2$, and $r_n\sim 0.3$, the equivalent of
two new neutrino families. This is unacceptable within the
standard model. A
model with an intermediate symmetry breaking scale at energy
$\lap H_x$ can give a  substantially larger $N_s$.
Alternatively, in supersymmetric theories there is a complex scalar field for 
each 
chiral quark or lepton. For the minimal model this gives
$N_s=104$ and $r_n\sim 0.01$, well within the bound in 
Eq. (\ref{35}).

The spectrum of gravitational waves differs
from the usual inflation picture \cite{Allen} 
because of the long kinetic energy-dominated period.
Gravitational waves are produced with 
strain per logarithmic interval of frequency $h_g \sim H_x/\mp$
at Hubble length crossing during inflation. 
The amplitude is nearly constant\footnote{More accurately, when
the proper wavelength is much larger than the Hubble length, and
$a\propto t^{1/3}$ during kinetic energy-dominated expansion, the
more rapidly growing solution to the field equation  
$\ddot h_g + 3\dot h_g\dot a/a = 0$ is $h_g=\ln t$ \cite{MG}. Here we ignore 
this slow
evolution.  A more careful analysis by Giovannini \cite{MG} shows
that it results in a logarithmic correction to the gravitational wave
spectrum (\ref{43}), $h_g \propto \omega^{-1/2}\ln (\omega_x /\omega)$.}  
until the proper wavelength re-crosses the Hubble length at expansion parameter
$a_\omega$, after which the amplitude decreases as $1/a(t)$ to
present value 
\beq
h_g \sim (H_x /\mp )a_\omega/a_o, \qquad H(a_\omega )\sim \omega a_o/a_\omega ,
\label{40}
\eeq
at present expansion parameter $a_o$ and present proper frequency
$\omega$. Waves that pass the Hubble
length at the time of equality of mass densities in relativistic
and nonrelativistic matter have present frequency 
\beq
\omega _{\rm eq } \sim 3\times 10^{-16}\Omega _m\hbox{ Hz}.
\label{41}
\eeq
Waves that pass the Hubble length at the end of inflation and at the start of 
radiation-dominated expansion have present frequencies 
\beqa
\omega _x&\sim & T_oR^{-1/4}\sim 10^{11}\hbox{ Hz}, \nonumber \\
\omega _r&\sim &\lambda R^{3/4}T_o\sim 10^{-3}\hbox{ Hz},
\label{42}
\eeqa
for our parameters. The part of the gravitational
wave spectrum that starts oscillating during kinetic energy-dominated
expansion, at 
$\omega _{r} < \omega < \omega _x$, is \cite{MG}
\beq
h_g\sim {T_o^{3/2}\over R^{3/8}\mp\omega ^{1/2}}
	\sim 10^{-24}\left(\omega _r\over\omega\right) ^{1/2}.
\label{43}
\eeq
Between $\omega _r$ and
$\omega _{\rm eq}$ the frequency 
dependence changes to $h_g\propto\omega ^{-1}$, the same as usual
inflation models. 

The form of the gravitational wave spectrum at 
$\omega >\omega _r$ can serve as an observational
signature of quintessential inflation. The signal in our example is too small to 
be 
detected by interferometric graviational wave detectors under construction, 
including 
LIGO and VIRGO, but may be within reach of future detectors. 

\section{Discussion}

The model meets several observational constraints. Primordial  density
fluctuations are essentially the same as in a chaotic inflation
model\cite{Linde}, and lead to the adiabatic cold dark matter model for
structure formation. This is viable but not yet shown to be
unique. Perhaps topological defects that originate near the end of inflation,
or the fields that produce the entropy in this model, also play a role
in structure formation. The expansion becomes radiation-dominated at
redshift $z\sim 10^{18}$, well before light element production, and
the energy density in gravitational waves may be
tolerable. Baryogenesis remains an open issue, as in conventional
inflation\cite{Joyce1}, \cite{Joyce2}. The model is adjusted to
present-day density parameter $\Omega _m=0.3$. There is considerable
evidence for this low value\cite{Peeb}. There is evidence for the zero
space curvature required here and in usual inflation, from the
redshift-magnitude relation for type Ia supernovae\cite{Saul},
\cite{Reiss}. This difficult measurement should be checked for
consistency with other constraints on $\Lambda$. The reading of
evidence from the anisotropy of the thermal cosmic background
radiation (the CBR) on open {\it vs.} cosmologically flat models still
is mixed\cite{Gaw}, \cite{Gor}.

Our model fails to show the tracking of the inflaton and matter mass densities
proposed by Zlatev, Wang and Steinhardt\cite{QUINT2}. We have checked that in 
the range of values $2 < \alpha < 6$ for the potential
$V\propto\phi ^{-\alpha }$ (Eq. \ref{5a}) well after inflation the energy 
density $\rho _\phi$ in the inflaton is quite close to 
constant over ten orders of magnitude of expansion centered 
on the present epoch. During this time the inflaton pressure is quite close to 
$-\rho _\phi$, so the field does not respond significantly to gravitational 
perturbations. 
Thus we suspect that, in contrast to the tracking 
case\cite{lensing},\cite{QUINT}, it will
 be difficult to find classical cosmological tests that distinguish our model 
from a Friedmann-Lema\^\i tre model with constant Lambda. 

There are arguably unnatural features of our model. First,
the model parameters have to be tuned to arrange inflaton domination
at the present epoch.  This problem is common to all cosmological
scenarios with $\Omega_m < 1$.  Second, it is somewhat unnatural for a
small mass $M\ll m_{pl}$ to appear in the potential of the inflaton
field $\phi$ interacting only with gravity.  Both problems can be
addressed in models with an extra inflaton field $\chi$ in which $M$
is not a constant but a slowly varying function of $\chi$.  Different
parts of the universe would then thermalize with different values of
$M$, and a low value of $M$ in our neighborhood could be selected by
anthropic considerations. Regions with $M \gg 10^6$ GeV would become
inflaton-dominated much earlier, and galaxy formation is such regions
would be suppressed.  This is similar to the anthropic selection of
the cosmological constant, as discussed in \cite{Anth}.

The quintessential inflation model simplifies the role of the inflaton by 
decoupling 
it from the matter. It remains to be seen whether this will aid the search for a 
believable physical basis for the inflaton. 

\section{Acknowledgments}

We are grateful for discussions with Massimo Giovannini, Alan Guth,
Guy Moore, Paul Steinhardt, and Michael Turner. 
This work was 
supported in part at the Princeton Institute for Advanced Study 
by the Alfred P. Sloan Foundation and at Tufts University by the
National Science Foundation.

\begin{figure}
\centerline{\psfig{file=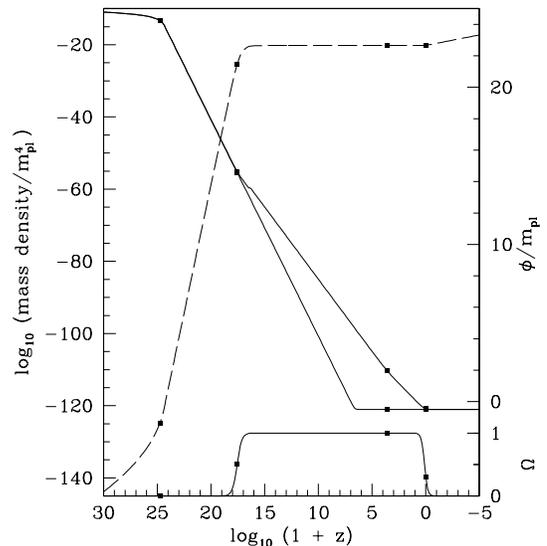,width=3truein,clip=}}
\caption{Evolution in quintessential inflation. The total mass density is 
plotted as a 
heavy solid line, the lighter line is the mass density in the inflaton, and the 
dashed 
line is the inflaton field value. The bottom curve is the density parameter (the 
fractional contribution to the square of the Hubble parameter by the mass 
densities in 
relativistic and nonrelativistic matter). 
The end of inflation, marked by the earliest (left-hand) 
filled squares, is defined by the maximum of $a(t)H(t)$ (the minimum value of 
the comoving 
Hubble length). Radiation-dominated expansion commences at the next squares. The 
next square 
indicates the epoch of equal mass densities in radiation and nonrelativistic 
matter. 
The last squares mark the present epoch at radiation temperature $T\sim 
3^\circ$~K. The 
parameter $M$ in the potential (Eqs.~\protect\ref{5} to \protect\ref{6a}) is 
chosen so 
the present value of the density parameter in matter is  
$\Omega _m=0.3$. 
The model for the matter assumes $R=1$ (Eq \ref{14}) and 
${\cal N}_{\rm th} = 1000$ effective scalar fields at first thermalization. 
Annihilation of the extra fields at $T=1000$~GeV causes the step in the mass 
density 
at redshift $z\sim 10^{16}$ (Eq. \ref{35}).} \end{figure} 

\end{document}